\documentclass[12pt,preprint]{aastex}
\def\({\left(}
\def\){\right)}
\def\[{\left[}
\def\]{\right]}

\usepackage{graphicx}
\usepackage{natbib}
\usepackage[usenames,dvips]{color}
\begin{document}
\title{Disk corona interaction: mechanism for the disk truncation and spectrum change in low luminosity AGN}
\author{Ronald E. Taam\altaffilmark{1,2}
\email{r-taam@northwestern.edu}
\and B. F. Liu\altaffilmark{3}, W. Yuan\altaffilmark{3}, \& E. Qiao\altaffilmark{3}
\email{bfliu@nao.cas.cn}
}
\altaffiltext{1}{Academia Sinica Institute of Astronomy and Astrophysics-TIARA,
P.O. Box 23-141, Taipei, 10617 Taiwan}
\altaffiltext{2}{Northwestern University, Department of Physics and Astronomy,
2131 Tech Drive, Evanston, IL 60208}
\altaffiltext{3}{National Astronomical Observatories, Chinese Academy of Sciences, 20A Datun Road, 
Chaoyang District, Beijing 100012, China}

\begin{abstract}
The truncation of an optically thick, geometrically thin accretion disk is investigated in the 
context of low luminosity AGN (LLAGN). We generalize the disk evaporation model used in the 
interpretative framework of black hole X-ray binaries by including the effect of a magnetic field 
in accretion disks surrounding supermassive black holes.  The critical transition mass accretion rate 
for which the disk is truncated is found to be insensitive to magnetic effects, but its inclusion 
leads to a smaller truncation radius in comparison to a model without its consideration.  That is, a 
thin viscous disk is truncated for LLAGN at an Eddington ratio less than 0.03 for a standard viscosity 
parameter ($\alpha = 0.3$). An increase of the viscosity parameter results in a higher critical 
transition mass accretion rate and a correspondingly smaller truncation distance, the latter 
accentuated by greater magnetic energy densities in the disk.  Based on these results, the 
truncation radii inferred from spectral fits of LLAGN published in the literature are consistent with
the disk evaporation model.  The infrared emission arising from the truncated geometrically thin accretion 
disks may be responsible for the red bump seen in such LLAGN.
\end{abstract}
\keywords{accretion, accretion disks --- black hole physics ---galaxies: active ---
X--rays:galaxies}

\section{Introduction}
The nature of accretion flows onto supermassive black holes (SMBHs) in active galactic nuclei (AGN) is of 
great interest in fundamental investigations of their emission properties and their growth through 
cosmic time. There is increasing support for the view that the accretion flow in the immediate vicinity 
of the black hole occurs via a geometrically thin, optically thick disk in the high/soft states of black 
hole X-ray binaries (BHXRBs) and via a geometrically thick, optically thin ADAF/RIAF in the low/hard states 
of these systems. By analogy, the dominant accretion flows in high luminosity AGN (HLAGN) are interpreted 
in terms of a thin disk while low luminosity AGN (LLAGN) are interpreted in terms of ADAFs. Evidence in 
support of a standard optically thick disk in HLAGN is often provided by the presence of a thermal optical/UV component,  and possibly extending to soft  X-rays (Yuan et al. 2010),  in radio quiet QSOs and some Seyfert galaxies.  On the other hand, the broad band emission 
spectrum from the SMBH in the Galactic Center has been successfully modeled in terms of an ADAF (Yuan et al.  2003). 

The observational data of LLAGN are commonly interpreted within a framework in which the emission arises
from accretion flows consisting of an inner ADAF connected to an outer geometrically thin disk.  That is, an ADAF 
with a surrounding optically thick disk is found to be necessary in modeling the broad band spectrum from 
the radio to X-ray energy regime of LLAGN  (Lasota et al. 1996; Quataert et al. 1999; Di Matteo et al. 
2000; 2003; Yuan et al. 2009; Li et al. 2009; Nemmen et al. 2006; 2012).  The need for this truncation 
derives from both the continuum and line emission and, at times, from their variability.  For example, the 
red bump in the continuum (e.g. Lasota et al. 2006; Ho 2008) can arise from the optically thick emission 
of an outer truncated disk.  In addition, evidence has been presented for a change in the accretion mode 
at an Eddington ratio of $\sim 0.01$ based on the break of the X-ray photon index (see Constantin et al. 
2009), as well as large-amplitude X-ray variability (Yuan et al. 2004). 
Recently,  Best \& Heckman (2012) investigated a large sample of radio-loud AGN and found that 
high-excitation sources typically have accretion rates between one per cent and 10 per cent of their 
Eddington rate, whereas low-excitation sources predominately accrete at a rate below one per cent 
Eddington, supporting the change of accretion mode at an Eddington ratio of 0.01 from a thin disk 
in the high-excitation sources to an ADAF in the low-excitation sources. We note that the 
high excitation radio loud sources may be analogous to the BHRXBs in the intermediate 
state, which have been successfully interpreted in terms of a transition from a truncated disk to 
a standard thin accretion disk (see Liu et al. 2006).
For the emission line diagnostics, the relation for the location, R, of the broad emission lines 
(e.g.$H_\beta$) with respect to luminosity, L, may deviate from the $R\propto L^{1/2}$ relation in some 
particularly dim AGN, which can also imply that the disk is truncated (Czerny et al. 2004; Liu \& Taam 2009).  
Furthermore, the iron line features in which the FWHM is connected to the Keplerian motion of gas in the 
inner radius of the disk suggest a larger radius than the innermost stable circular orbit (ISCO) (e.g. NGC 
4593 in Markowitz \& Reeves 2009). Finally, the double-peaked Balmer emission lines, in particular,  
the rarity and shapes of their profiles, have been speculated as  
arising from a flat, rotating, outer/truncated disk observed at a large inclination (see e.g., Storchi-Bergmann et al. 2003; Wang et al. 2005; Eracleous, Lewis \& Flohic 
2009).  Although support for such a picture has been provided by a number of phenomenological studies, an understanding of disk truncation and the spectral transition to an ADAF at low mass accretion rates
remains a challenge.  Early work by Wandel and Liang (1991) demonstrated the existence of multiple
solutions in a hybrid disk model, which included a hot inner disk that could, in principle, be
connected to a cool outer disk for a viscous stress proportional to gas pressure (see also Hubeny
1990).  However, the preference for this solution over a purely optically thick gas pressure solution
was not addressed and was only hypothesized by Narayan \& Yi (1995) in the "strong ADAF principle"
(if an ADAF can form, it will).

Mechanisms to facilitate the truncation of a thin disk at low accretion rates have been investigated by 
a number of groups including Honma (1996), Manmoto \& Kato (2000), Lu, Lin, \& Gu (2004), Meyer, Liu, 
\& Meyer-Hofmeister (2000a,b), R\`o\.za\`nska \& Czerny (2000a,b), Spruit \& Deufel (2002), and Dullemond 
\& Spruit (2005). Among the proposed models (see Table 1), the disk evaporation model based on vertical thermal conduction (Meyer \& Meyer-Hofmeister 1994; 
Meyer et al. 2000a, 2000b; Liu et al. 1999, 2001, 2002,2005,2006) has been most extensively investigated.
It naturally provides an explanation for the truncation of the thin disk in the low/hard state observed in BHXRBs.

\begin{table}[htdp]
\caption{Models proposed for disk truncation}
\begin{center}
\begin{tabular}{|l|l|}
\hline  Model & References\\
\hline Strong ADAF principle & Narayan \& Yi 1995\\
\hline Disk evaporation & Meyer, Liu, \& Meyer-Hofmeister 2000a,b; R\`o\.za\`nska \& Czerny 2000a,b;\\
&  Spruit \& Deufel 2002, and Dullemond  \& Spruit 2005\\
\hline Radial conduction & Honma 1996; Manmoto \& Kato 2000; Gracia et al. 2003 \\
\hline Thermal instability & Takeuchi \& Mineshige 1998; Gu \& Lu 2000; Lu, Lin, \& Gu 2004\\
\hline
\end{tabular}
\end{center}
\label{default}
\end{table}%

The disk evaporation model has been applied to AGN in an earlier study (Liu \& Taam 2009), where the 
detailed dynamics and equations were described and the evaporation features as a function of accretion 
rate were presented.  As a consequence of the complete evaporation of the inner optically thick regions of 
the disk at low accretion rates, a spectral state transition of AGN was predicted and the possible 
disappearance of the broad line region at low luminosities was discussed.  In this work, we specifically 
focus on LLAGN where the accretion rate is low, for which the disk can be truncated. Here, 
the dependence of 
coronal flow and truncation radius on the accretion rate is investigated taking into acccount the effects 
of viscosity and magnetic fields. The numerical results of our models for evaporation and disk truncation 
are approximated by analytical fits in Section 2.  Based on the determination of the truncation radius 
from our models, typical spectra from an inner ADAF and truncated disk are presented for LLAGN in Section 
3. The predictions of the disk truncation model are qualitatively compared with the published spectral 
fits to observed data of individual LLAGN in Section 4.  We discuss further support for the disk 
evaporation model and the observational inferences that can be used to provide constraints on the viscosity 
and magnetic fields in the disk in Section 5. Finally, we conclude in the last section. 

\section{The Disk Corona Model}
The disk evaporation model for accreting black holes was first established in detail by Meyer, Liu, 
\& Meyer-Hofmeister (2000b), based on the pioneering work of Meyer \& Meyer-Hofmeister (1994) on 
dwarf novae. The model was further developed to include the decoupling of ions and electrons (Liu et al.
2002) and the inflow and outflow of mass, energy, and angular momentum between neighboring zones 
(Meyer-Hofmeister \& Meyer 2003).  Detailed calculations incorporating the influence of a magnetic field and 
viscosity was subsequently studied by Qian, Liu, \& Wu (2007) and Qiao \& Liu (2009).  In this model, a hot 
geometrically thick, optically thin corona is formed above a geometrically thin standard disk by
processes similar to those operating in the surface of the Sun, or by a thermal instability in the 
uppermost layers of the disk (e.g., Shaviv \& Wehrse 1986). Both the disk and corona are individually 
powered by the release of gravitational energy associated with the accretion of matter affected through
viscous stresses.  In the corona, the viscous heat is partially transferred to the electrons by means of 
Coulomb collisions. The energy is conducted into the lower, cooler, and denser layer and radiated away 
in the chromosphere.  If the density in the lower corona is too low to efficiently radiate this energy,
a suitable amount of cool matter is heated and evaporates into the corona. This mass evaporation continues 
until an equilibrium density is established.  By frictional processes, the gas loses angular momentum and 
accretes through the corona to the black hole.  The accretion flow is maintained by a steady mass 
evaporation flow from the underlying cool disk.
 
 The disk corona is physically similar to an ADAF except that vertical conduction inevitably occurs from the corona to the underlying, much cooler disk, which cools the corona more efficiently than radiation. We  assume an $\alpha$ prescription for the viscosity and an equipartition parameter $\beta$, the ratio of gas pressure to the total pressure, for the magnetic field.  There could be physical connection between the viscosity and magnetic fields, however, we take $\alpha$  and  $\beta$  as independent since the 
magnetorotational instability does not provides a sufficiently large  $\alpha$ (see King et al. 2007).  
  The model used here is based on Liu et al. (2002) and includes modifications (Meyer-Hofmeister \& Meyer 
2003) and updates incorporated in recent years (e.g., Qian et al. 2007; Qiao \& Liu 2009). The structure 
of the corona and evaporation features are determined by the equation of state, equation of continuity,
and equations of momentum and energy (for details see Liu \& Taam 2009).
Numerical calculations show that the evaporation increases with decreasing distance in the outer region
of the disk, reaching a maximum and then decreasing toward the central black hole. In the innermost region 
the evaporation can be replaced by condensation provided a small optically thick cool disk is present (see 
Meyer, Liu, \& Meyer-Hofmeister 2007; Liu et al. 2007; Taam et al. 2008). Thus, the mass flow rate in the corona supplied by the mass evaporation 
increases towards the central black hole unless condensation occurs.  Therefore, for a given mass supply 
rate to the disk ($\dot m$), evaporation diverts an increasing flow to the corona with decreasing distance.
The remaining part of the mass flow in the disk decreases as
\begin{equation}
\dot m_d(r)=\dot m-\dot m_{\rm evap}(r),
\end{equation}
where the rates are in units of the Eddington accretion rate ($\dot M_{\rm Edd}= 1.39\times 10^{18} 
M/M_\odot\,{\rm g\,s^{-1}}$), and $\dot m_{\rm evap}(r)$ is the sum of the evaporation rate from the outer
edge to distance $r$ (in units of the Schwarzschild radius $R_s=2GM/c^2$).

For accretion rates higher than the maximum evaporation rate, the evaporation can only divert a fraction of 
the disk accretion flow to the corona and the optically thick disk is never completely truncated by evaporation. 
However, for an insufficient mass supply to the disk, $\dot m< \dot m_{\rm evap}$, the gas in the inner regions of 
the disk can be fully evaporated to form an optically thin region, leading to disk truncation. The truncation 
radius is determined by,
\begin{equation}
\dot m-\dot m_{\rm evap} (r_{\rm tr})=0.
\end{equation}

Detailed investigations show that the mass evaporation rate is a function of the viscosity parameter, 
$\alpha$, and magnetic field (parameterized by the ratio of gas pressure to the total pressure), $\beta$. 
An increase in $\alpha$ leads to an efficient heating of the corona. In the inner region, this viscous heat is partially transferred to electrons through Coulomb collisions and conducted down to the transition layer, 
resulting in an increase in the mass evaporation rate. On the other hand, the gas density in the 
outer corona is very low and little heating of the electrons occurs through Coulomb collisions 
leading to almost no effect on the evaporation rate in the outer corona. 
Therefore, an increase of $\alpha$ leads to an increase of mass evaporation rate in inner corona.  
The effect of magnetic fields is a competition between its tendency to increase the evaporation as a 
result of energy balance and to decrease the evaporation as a result of pressure balance.  The magnetic field contributes additional pressure to the corona, which results in more heating via the shear stress. The effect is similar to an increase in $\alpha$,  it leads to an increase of evaporation rate in the inner region while little effect in the outer region.  On the other hand,  the additional 
pressure contribution inhibits the evaporation as a result of pressure balance. The combined effect of 
the magnetic field, as shown from our numerical calculation, reveals almost no change in the 
maximal evaporation rate, but with a reduction of the evaporation rate at a given distance in the 
outer region. 

Therefore, for a steady mass supply/accretion  rate to the disk,  
the variation of the truncation radius as a function of mass accretion rate is illustrated in Fig.1
for several values $\alpha$ and $\beta$.  Specifically, the maximum evaporation rate and its 
corresponding radius vary with $\alpha$ approximately as (Qiao \& Liu  2009)
\begin{equation}\label{tr-alpha}
\dot m_{\rm max}\approx 0.38\alpha^{2.34}\
{\rm and}\  r_{\rm min}\approx 18.80\alpha^{-2.00},
\end{equation}
which is derived from numerical results for $0.1\la \alpha\la 0.9$.

The dependence of the maximum 
evaporation rate and the corresponding truncation radius on magnetic fields  for given viscosity $\alpha=0.3$ can be approximated by (Qian, Liu, \& Wu 2007) 
\begin{equation}\label{tr-beta}
\dot m_{\rm max}\approx 0.026\beta^{-0.41}\  {\rm and} \  r_{\rm min}\approx 209\beta^{4.97},
\end{equation}
 which is derived from fits to numerical results for $ 2/3<\beta<1$ and assumed valid for magnetic fields constrained by equipartition, $ 0.5<\beta<1$. 
It can be seen that the maximum evaporation rate varies slightly with the magnetic field strength, whereas the truncation of  the disk occurs at much smaller radii for smaller $\beta$ (see Fig. 1). 

For $\alpha=0.3$ and $\beta=1$, an approximate fit to the evaporation curve of the outer part 
of the disk at large truncation radii (see Fig. 1) yields an expression for the truncation radius 
given by
\begin{equation}\label{tr-specific}
r_{\rm tr}\approx 15.9 \dot m^{-0.886}.
\end{equation}
Assuming that the shape of evaporation curve $\log \dot m_{\rm evap}-\log r$ is independent of the strength 
of the magnetic field and viscosity, which is borne out from the computational data (that is, 
the evaporation rate $\dot m_{\rm evap}$ at distance $r$ depends on the viscosity and magnetic field in a 
similar way as the maximum evaporation rate), 
the truncation radius can be generalized as
\begin{equation}\label{tr-radius}
r_{\rm tr}\approx 17.3\dot m^{-0.886}\alpha^{0.07}\beta^{4.61}.
\end{equation}
This functional dependence for the truncation radius is a good fit to the numerical data provided that the 
accretion rate is less than half of the maximum accretion rate, $\dot m\la{1\over 2}\dot m_{\rm max}
\approx 0.19\alpha^{2.34}$, corresponding to $\dot m\la 0.01$ for the standard viscosity $\alpha=0.3$. 
It can be seen that the truncation radius approximated by Eq.(\ref{tr-radius}) is weakly dependent on 
$\alpha$.  That is, the viscous parameter little affects the truncation radius for values near 0.3 in a 
hot accretion flow.

However, at accretion rates close to the maximum evaporation rate, the evaporation curve deviates from
the power-law fit (see Fig.1) of Eq. (\ref{tr-radius}). In this case, a transition from an ADAF to a thin 
disk is expected to take place and the truncation radius is determined from the numerical results. 
Taking into account the effects of viscosity and magnetic fields, we obtain from our numerical calculations 
that the maximum evaporation rate and corresponding truncation radius as a function of the viscosity and 
magnetic parameters are given as
\begin{eqnarray}
\dot m_{\rm max} \approx 0.38\alpha^{2.34} \beta ^{-0.41} \label{mdot-max}\\
r_{\rm min} \approx 18.80\alpha^{-2.00}\beta^{4.97}\label{tr-min}.
\end{eqnarray}
This reveals that the truncation radius 
can be a few tens of Schwarzschild radii if the viscous parameter is not very small and the magnetic fields 
are sufficiently strong.

It can also be seen in Fig. 1 that the accretion rate for a spectral state transition depends on the 
viscosity (solid lines).  For instance, at a mass accretion rate of 0.01, a transition from hard to soft 
state occurs for $\alpha=0.2$. On the other hand, the system remains in the low/hard state with the disk 
truncated at a distance of $\sim1000$ Schwarzschild radii for a larger viscosity.  Although the truncation 
distance is little affected by viscosity at low accretion rates, the disk can extend to $\sim 100$ 
Schwarzschild radii at high accretion rates before a state transition in the case of a larger viscosity. 
The magnetic field accentuates the truncation to small distances as shown by the dashed lines in Fig. 1.

\section{Model Predictions to the Spectral Energy Distribution of LLAGN}

The disk evaporation model results in a two component structure of the accretion disk composed of an 
inner optically thin region and outer optically thick region.  In this section, we examine the emission 
properties of each component to quantify their spectral characteristics. 

\subsection{Emission from the Truncated Disk}

The radial distribution of the effective temperature in a standard disk is determined by the black hole mass, $M$,  mass accretion rate, $\dot M$, and  distance, $R$,  
$\sigma T_{\rm eff}^4={3GM\dot M\over 8\pi R^3}\[1-(R_*/R)^{1/2}\]$, which for convenience in application to LLAGN we express as 
\begin{equation}\label{Teff-trun}
T_{\rm eff}(R)=6.237\times 10^5 r^{-3/4}\[{1-\({r_*\over r}\)^{1/2}}\]^{1/4}
\({m\over 10^8 }\)^{-1/4}\dot m^{1/4}K.
\end{equation}
This temperature reaches a maximum at $r=(49/36)r_*$. Assuming a non-rotating black hole, $r_*=3$, 
and the maximum effective temperature is
\begin{equation}\label{Teff-max}
T_{\rm eff, max}=1.335\times10^5 \({m\over 10^8}\)^{-1/4}\dot m^{1/4}K.
\end{equation}
For high accretion rates in the range of $\sim 0.1-1 \dot M_{Edd}$, characteristic of values inferred for 
HLAGN, the thin disk is not truncated according to our model. If the SMBH masses lie in the range of 
$10^6M_\odot-10^9M_\odot$, the maximum effective temperature of the disk (calculated from Eq.\ref{Teff-max}) 
is in the range between $4.2 \times 10^4$ K $- 4.2 \times 10^5$ K, producing radiation in the optical-UV or 
EUV energy band depending on the black hole mass and accretion rate.

However, at accretion rates lower than the maximum evaporation rate ($\dot M<0.027\dot M_{\rm Edd}$  for
$\alpha=0.3$ and in the absence of magnetic field effects) the accretion flows are characterized by disk 
truncation and the inner edge of the disk recedes to a distance greater than $200R_s$. In this case, the 
effective temperature in a truncated disk is less than 4600K for a $10^8M_\odot$ black hole (calculated from 
Eq.(\ref{Teff-trun})).  This temperature decreases with decreasing accretion rate not only due to the 
explicit $\dot M$ dependence, but also because the truncation radius shifts outwards. 

On the other hand, the truncation radius can be as small as a few tens of Schwarzschild radii taking 
magnetic field effects into account. This effect is significant since radiation from the truncated disk 
can peak at optical wavelengths in LLAGN.   Hence, the disk emission in LLAGN can peak in the infrared 
to optical wavelength regime, depending on the rate of mass accretion and the magnetic $\beta$ parameter. 

The luminosity of the truncated outer disk can be estimated from $L_d=0.1\dot M c^2 \times {3\over 
r_{\rm tr}}$ and is given by 
\begin{equation}\label{e:Ld}
L_{\rm disk}=0.1\dot M c^2 \[0.066\({\dot m\over 0.01}\)^{0.886}\({\alpha\over 0.3}\)^{-0.07}
\({\beta\over 0.5}\)^{-4.61}\]
\end{equation}
For comparison, the luminosity produced in the inner ADAF region of the disk can be approximated as 
(see Mahadevan 1997) 
\begin{equation}\label{e:LADAF}
L_{\rm ADAF}\approx 0.1\dot M c^2 \[0.20\({\dot m\over 0.01}\)\({g(T_e)\over 4}\)\({\alpha\over 0.3}\)^{-2}\({\beta\over 0.5}\)\],
\end{equation}
where $g(T_e)$ is in the range of $1.4\le g(T_e)\le 12$ for $10^9\le T_e\le 5\times 10^{9}$ and  $\alpha$ is  changed from the definition $\nu=\alpha c_s H$ in Mahadevan (1997) to our definition  equivalent to $\nu={2\over 3}\alpha c_s H$.

Given the luminosities from the ADAF and the optically thick disk, we find that the disk luminosity is 
less than the  ADAF luminosity even in disks characterized by magnetic energy densities in equipartition 
with the plasma energy densities (corresponding to a small truncation radius). 
 Hence, for the case of weak magnetic fields, the emission 
from the ADAF is dominant.  We point out that this conclusion is unchanged in the RIAF model where direct 
heating (characterized by a parameter $\delta$) of the electrons is assumed to arise from the viscous 
dissipation. This leads  to an additional component to the ADAF luminosity given as, 
\begin{equation}\label{e:LADAF2}
L_{\rm ADAF}\approx 0.1\dot M c^2 \[0.40\({\delta\over 0.5}\)\({1-\beta\over 0.5}\)\({f\over 0.5}\)^{-1}\],
\end{equation}
where $f$  is the advected fraction of the viscously dissipated  energy (Narayan \& Yi 1995). This 
contribution to the luminosity from the ADAF can exceed that due to the Coulomb collisional transfer 
process  if $\delta$ is much larger than the mass ratio of electron to proton,  and, thus, can dominate the ADAF luminosity for low mass accretion rates, i.e., $\dot m< 0.02\({\alpha\over 0.3}\)^2 \({1-\beta \over 0.5}\)\({\beta\over 0.5}\)^{-1}\({\delta\over 0.5}\)\({c_1\over 0.5}\)^2 \({f\over 0.5}\)^{-1}\({g(T_e)\over 4}\)^{-1}$.
On the other hand, wind loss in the inner ADAF flow ($\dot m(r<r_{\rm tr})\propto r^{s})$ can counter 
balance this effect as it results in a decrease of the luminosity from the ADAF.  These two effects 
partially cancel to some extent and do not significantly affect the strength of disk component relative 
to the ADAF, $L_{\rm disk}/L_{\rm ADAF}$, as calculated from Eqs.(\ref{e:Ld}) and (\ref{e:LADAF}).

\subsection{Spectral Energy Distribution from ADAF and Truncated Disk} 

Given the mass of the SMBH and the mass accretion rate, the truncation radius is determined from the disk 
evaporation model for a given value of $\alpha$ and $\beta$.  
The radiation from the inner ADAF and the outer disk and its resulting spectral energy distribution (SED) 
is calculated (for details see Qiao \& Liu 2010). Specifically, we describe the inner ADAF/RIAF by the 
self-similar solution as a function of  $m$ and $\dot m$ (Narayan \& 
Yi 1995)  with fixed $\alpha=0.3, \beta=2/3$.  The emissions in the ADAF include contributions from bremsstrahlung, synchrotron radiation 
and the Comptonization of the bremsstrahlung, synchrotron radiation and the soft photons emitted by the 
truncated disk.  The outer region is described by a two-phase disk corona accretion flow.  The 
truncated disk and corona structure, characterized by the electron temperature and density distribution, 
are determined by the disk evaporation model.  Multi-color blackbody radiation from the truncated disk is 
included in the emergent spectra.  We consider models characterized by $m=10^{8}$, and $\alpha=0.3$ for 
mass accretion rates of $\dot m = 0.01$ and 0.02 and $\beta = 2/3$ and 1. For these cases, the truncation 
radius, as obtained from our detailed structure calculations, is $r_{\rm tr}=107$ ($\dot m = 0.02$, $\beta 
= 2/3$), $r_{\rm tr}=236$ ($\dot m = 0.01$, $\beta = 2/3$), and $r_{\rm tr}=1000$ ($\dot m = 0.01$, $\beta 
= 1$). The resulting SEDs are plotted in Fig.\ref{f:spectra}.  It can be seen that two peaks exist in the 
spectrum, one at lower energies corresponding to the optically thick disk component and the other at higher 
energies corresponding to the ADAF component.  The contribution of the disk to the bolometric luminosity is 
lower than the ADAF contribution even at an accretion rate of 0.02,  indicating that the  spectra at 
accretion rates lower than 0.02 are typical hard-state spectra.  It is clear that an increase in the field 
(decrease in $\beta$) as well as an increase in the mass accretion rate shifts the peak of the spectrum 
arising from the disk component to shorter wavelengths from $\sim 10 \mu m$ to $\sim 1.6\mu m$. 
If the magnetic field is stronger than the cases as shown in Fig. 2, the disk can be truncated at a few 
tens of Schwarzschild radius and its blackbody radiation can reach the optical waveband.  A color correction 
could further shift the disk component toward shorter wavelengths (Shimura \& Takahara, 1995; Davis et al. 2005).  
However, Done et al. (2012) have recently shown that for temperatures less than $3 \times 10^4$ K, as in 
the low/hard state AGN, the importance of absorption relative to scattering in the disk obviates the need for 
such a correction. This suggests that magnetic fields can be important in LLAGN in which an optical 
blackbody component is observed. 
 
 We note that a disk component  as shown in Figure 2 is observed in a number of LLAGN as 
shown in the next section. 

\section{Spectral Fits to Observational SEDs of LLAGN}
In fits to the observational spectra of a LLAGN, an ADAF and an outer thin disk are commonly assumed 
with a truncation radius between the disk and ADAF taken as a free parameter without relevance to the Eddington 
ratio. In addition, the ADAF model, itself, is characterized by several parameters, viz., a fraction 
($\delta$) of viscously dissipated energy is assumed to directly heat the electrons (see above) and 
wind loss which leads to the variation of the mass flow rate as a function of the radius.  This wind loss 
leads to a decrease in the mass flow rate as the SMBH is approached and is described by a power law in the 
form $\dot m (r<r_{\rm tr})\propto r^{s}$. Hence, $\delta$ and $s$  are free parameters in addition to the 
accretion rate and truncation radius used in modeling the spectrum from the ADAF/RIAF. As a result, there 
are often more than one set of parameters  which can fit the spectrum equally well, i.e., the values 
of $\dot m$ and $r_{\rm tr}$ change for different $s$ and $\delta$.  Taking into account this degree of 
freedom in the fitting, we deem it unnecessary to repeat the fits to individual sources. However, to reduce 
the degeneracy of the solutions, we make use of the disk evaporation model to constrain the truncation 
radius from the available fitting results.  In this way, a fit to the observational SED can be adopted for 
consistency within the framework of our theoretical model.

Spectral fits based on an ADAF/RIAF and a truncated disk model for individual LLAGN have been considered 
ever since the ADAF model was established. For example, Quataert et al. (1999) modeled the optical/UV 
spectrum of M81 for a mass accretion rate of $\dot m=0.01$ and an inner radius of the optically thick 
disk of $r_{\rm tr}=100$. Within the framework of the disk evaporation model, this radius requires the need 
for magnetic effects and amounts to $\beta=0.61$ based on Eq.(\ref{tr-radius}). The viscosity parameter is 
approximately the standard value, for which the truncation radius is insensitive.  Similarly, a fit to the 
optical/UV spectrum of NGC 4579 yields $\dot m=0.03$ and $r_{\rm tr}=100$ (Quataert et al. 1999).  Assuming 
$\beta=0.76$ in the corona, the truncation of the thin disk at $r_{\rm tr}=100$ can be interpreted as a 
result of disk evaporation.  We note that the accretion rates for these two objects are close to the 
maximum evaporation rate 0.027 for the case corresponding to the standard viscosity parameter ($\alpha=
0.3$) in the absence of magnetic effects. Combining this with the relatively small truncation radius at 
$100R_s$, it is possible that these objects are at a stage close to a state transition. If this is the 
case, the disk truncation features for these two objects can be fitted as well by taking $\alpha=0.2$ 
and $\beta=0.73$ for M81, and $\alpha=0.33$ and $\beta=0.89$ for NGC4579 (from Eqs. (\ref{mdot-max}) 
and (\ref{tr-min})).

Recently Xu \& Cao (2009), using updated SMBH mass estimates, have obtained fits to these two objects and 
showed that the Eddington scaled accretion rate can be an order of magnitude smaller than that inferred by 
Quataert et al.  (1999), while the truncation radii are similar.  For such a low accretion rate, $\dot m\la10^{-3}$, 
only very strong magnetic fields ($\beta \sim 0.4$) can result in truncation at $r_{\rm tr}\sim 100$ in the model. 

Additional sources were examined by Di Matteo et al. (2000), who modeled the spectra of 6 elliptical galaxy
nuclei using ADAFs truncated within $10^4$ Schwarzschild radii. In this study, the mass accretion rate is 
assumed to decrease as a power law from the truncation radius toward the SMBH and the power-law index was 
taken as a free parameter in addition to the truncation radius and mass accretion rate. The fits indicated
that the truncation radii  and accretion rates lie in a large range ($100\la r\la 10^4$,  
$\dot m\sim 10^{-3}$ for the case of no wind and $\dot m \sim 0.01-0.03$ with a strong wind).  Nevertheless, 
the derived truncation radii and accretion rates fall within the range of our model 
prediction, though there exist large uncertainties in the fitting parameters.  Thus, the disk truncation 
in these objects can also be ascribed to disk evaporation with the effect of the magnetic fields playing 
an important role in sources characterized by smaller truncation radii.
  
The accretion mode of an inner ADAF/RIAF connecting to a truncated disk has also been proposed as the 
source of emission in models for X-ray bright, optically normal galaxies (Yuan \& Narayan 2004).  In 
particular, fits to the Chandra and XMM-Newton data and optical/UV emissions for two such sources (XMM 
J021822.3-050615.7 and P3) reveal that a thin disk is truncated at a distance of $r_{\rm tr}=60$ with an
accretion rate $\dot m=0.01$ (Yuan \& Narayan 2004).  The truncation can be interpreted as a consequence 
of disk evaporation by adopting $\beta=0.55$ (Eq.\ref{tr-radius}). Such a $\beta$ value indicates the 
existence of a strong magnetic field close to equipartition.  This would be consistent with 
the assumption in the RIAF modeling that half of the viscous dissipation directly heats the electrons, 
which could be facilitated by magnetic instabilities.  If these sources are at the transition stage, the 
viscosity in these sources could be slightly smaller, with $\alpha=0.2$, and the required magnetic fields 
slightly weaker,  $\beta=0.66$ (from Eqs. (\ref{mdot-max}) and (\ref{tr-min})).

In a very recent study, Nemmen et al. (2012) compiled a sample of 21 SEDs of LLAGN found in LINERs which 
included observations of high spatial resolution in the optical and UV energy bands obtained with the 
Hubble Space Telescope, X-ray observations with Chandra, high-resolution radio observations with the 
Very Large Array or VLBA/VLBI, and available infrared observations for sources with SMBH mass 
determinations.  The radio to X-ray SEDs of the 21 LLAGN were modeled using three components: an inner 
ADAF/RIAF, an outer truncated geometrically thin accretion disk and a jet. It is found that the X-rays 
can be fit by either the jet or the ADAF for most of the sources,  while the radio band is almost always 
dominated by the synchrotron emission from the jet.  The near IR to optical band is found to be dominated 
by the truncated thin disk thermal emission.  For NGC 1097 (see also Nemmen et al. 2006), NGC 4143, NGC 
4278 and NGC 4736, a truncated thin disk is required to fit the mid- and near-IR data,  with the resulting 
transition radii in the range $30-225 R_s$.  

In the ADAF/RIAF and truncated thin disk model, the accretion rates resulting from fits to the entire sample 
of Nemmen et al. (2012) are mainly in the range of 0.003 to 0.01 and the truncation radii are in the range 
of $40-10^4 R_s$.  The truncation of the disk can be interpreted as due to gas 
evaporation if an appropriate magnetic field strength is adopted.  A large number of the sources show 
an accretion rate $\sim 10^{-3}$ and truncation radius of $\sim 10^{4}$,  which are in good agreement 
with the prediction of the disk evaporation model (Eq.\ref{tr-radius}) for a standard viscosity parameter in 
the absence of a magnetic field ($\alpha=0.3, \beta=1$). Sources with  relatively small truncation radii 
of $\sim 100R_s$  derived from SED fits required strong magnetic fields.  For example, in the case of NGC 
1097, the accretion rate and truncation radius derived from the fits to the SED are $\dot m=6.4\times 
10^{-3}$ and $r_{\rm tr}=225$ respectively.  The disk evaporation model predicts a consistent truncation 
radius at an accretion rate of $\dot m=6.4\times 10^{-3}$ if $\beta=0.67$ is adopted.  Although the 
deduced value of $\beta$ in the corona may differ from the adopted one in the ADAF/RIAF model in fits to 
the SED, it does not change our conclusion since the 
X-ray spectral slope from an ADAF depends only weakly on the value of $\beta$ (and also $\alpha$).  Hence, the SED is 
equally well fit for a different value of $\beta$. 

It should be pointed out that most of the sources in the sample of Nemmen et al. (2012) were previously 
studied using an ADAF and truncated disk model. In the following, we summarize the results of the detailed 
spectral fitting for a few sources (M81 and M87) to illustrate the qualitative agreement with the disk 
evaporation model and the constraints on $\alpha$ and $\beta$.  The SED of M87 was modeled by Di Matteo et al. 
(2003) based on an ADAF model with various values of $\delta$, but without wind loss (s = 0). 
Although Yuan, Yu \& Ho (2009) found that a jet model led to a better fit than the RIAF if $\delta$ 
is 0.5 as required by fits to Sgr A* (Yuan et al. 2003), the recent model of Nemmen et al. (2012) (similar to 
that of Di Matteo et al. 2003), results in a fit with an accretion rate of $\dot m=5.5\times 10^{-4}$ and 
truncation radius of $r_{\rm tr}= 10^{4}$, with little direct heating to electrons ($\delta=0.01$) and wind 
loss ($s=0.1$) required. We note 
that these results are in good agreement with the expectations from the disk evaporation model for the 
standard viscosity parameter, $\alpha\approx 0.3$ and $\beta\approx1$, as predicted by Eq.(\ref{tr-radius}). 
 
M81 is also well studied, having abundant multi-wavelength observational data.  The $H_\alpha$ line profile of 
M81 provides a diagnostic probe of the inner optically thick disk as Devereux \& Shearer (2007) found it to be 
consistent with an inner radius of $280-360R_s$ for the thin disk. This is in contrast to the result of 
Young et al.  (2007), who favored a smaller transition radius in modeling the Fe $K_\alpha$ emission line.  
As described above, the SED was fit by Quataert et al. (1999) with an ADAF and a truncated disk model assuming 
$\dot m=0.01$, $r_{\rm tr}=100$, $\delta=0.01$ without the inclusion of winds or jets.  This compares with 
the work of Nemmen et al. (2012), who found a lower accretion rate and larger truncation radius, $\dot m=
0.003$, $r_{\rm tr}=360$ using $\delta=0.01$ and $s=0.16$. Given that the truncation radius is constrained 
by the observation of $H_\alpha$ lines, the truncation at $r_{\rm tr}=360$ at an accretion rate of $\dot m=
0.003$ requires a strong magnetic field ($\beta=0.65$), which is slightly stronger than that inferred from the 
fits obtained by Quataert et al. (1999).  We note that there is a degeneracy in this fitting since the SED of 
M81 can also be fit by a jet model under different assumptions (see Yuan Yu, \& Ho 2009; Nemmen et al. 2012), 
in which it is found that the truncation radius, $r_{\rm tr}=50$,  is closer to the estimate from the Fe 
$K_\alpha$ line profile. 
  
Among the remaining sources reported in Nemmen et al. (2012), NGC 3998 and NGC 4278 can be quantitatively 
examined within the framework of the disk evaporation model.  Specificially, the SED of NGC 3998 was 
modeled by Ptak et al. (2004) and subsequently confirmed by Nemmen et al. (2012). Both the ADAF and truncated 
disk model and the jet model can reproduce the broadband SED. The fitting parameters in the former model 
correspond to  $\dot m=7.2\times 10^{-3}$ and $r_{\rm tr}=10^4$ when a strong wind loss was assumed, $s=0.4$. 
The truncation radius is not unique for this object as the SED can also be reasonably fit with a 
truncation radius  as small as $500 R_s$ for  an accretion rate of $10^{-3}$ in models for which a lower 
wind loss is assumed ($s=0.01$ for $r_{\rm tr }=500$ and $\dot m =10^{-3}$). However, it cannot be much 
smaller than $500R_s$, otherwise the emission of the truncated thin disk would exceed the IR upper limits. 
The lack of Fe $K_\alpha$ line emission (Ptak et al.  2004) also suggests that the value of $r_{\rm tr}$ is 
not so small. The disk evaporation model can interpret  the truncation from 500 up to a few thousand 
Schwarzschild radii for $\dot m\sim 10^{-3}$.
A strong magnetic field ($\beta\sim 0.56$) is required  to ensure that the 
optically thick disk is truncated at a distance of $500 R_s$.  

NGC 4278 is also an interesting object in this context as its radio emission was investigated within the 
framework of ADAF models by Di Matteo, Carilli \& Fabian (2001), who found that $s > 0$ is required.  
Nemmen et al. (2012) showed that both the ADAF and jet model require a truncated thin disk with a small 
transition radius ($r_{\rm tr} \sim 30-40$) in order to account for the near-IR data.  However, the conclusion 
that the truncation radius is small is not unique.  A larger radius,  $r_{\rm tr}=100$, can also reproduce 
the SED, but for a higher mass accretion rate, $\dot m=4\times 10^{-3}$, and for a larger wind loss, $s=0.77$. 
The disk evaporation model can be invoked to provide support for truncation at $100 R_s$, but fails to explain 
a very small truncation radius at an accretion rate as low as $\dot m=7\times 10^{-4}$.  Hence, we favor the 
solution where the optically thick disk is truncated at $\sim 100 R_s$ for this object.
      
Although a number of additional sources provide some evidence for disk truncation (e.g., Chiang \& 
Blaes 2003 for NGC 5548; Nemmen et al. 2012 and references therein), there are large uncertainties associated
with the determination of the bolometric luminosity and the SMBH mass.  As a result, there is greater freedom 
in deriving the mass accretion rate and truncation radius using an ADAF or RIAF model.  Hence, we defer the 
comparison of the disk evaporation model with these additional sources until the observational data 
are sufficient to provide better constraints on these fundamentally important quantities.

\section{Discussion}
\subsection{A Thin Disk as an Alternative Model for LLAGN?}
The model of an optically thick disk with an inner ADAF for the central engines of LLAGN was challenged 
by Maoz (2007) based on the property that LLAGN in LINERs are characterized by UV/X-ray luminosity ratios 
similar, on average, to those of brighter Seyfert Type 1 galaxies. Maoz (2007) pointed out that a geometrically 
thin disk extending to the ISCO could apply to both HLAGN and LLAGN, despite the 
much lower mass accretion rates inferred for LLAGN. On the other hand, Yu, Yuan \& Ho (2011) recently 
demonstrated that the SEDs of LLAGN compiled by Maoz (2007) can also be fit by an ADAF and truncated thin 
disk model when account is taken of the contribution from a jet. Although the single component 
thin disk model cannot be excluded for the origin of UV emission in some LLAGN, the radio and X-ray fluxes
require an additional component(s).  The comparisons of the spectral fits with an ADAF and truncated disk 
within the disk evaporation paradigm (see section 4) provide further support for the multi component 
picture that the thin disk is truncated and the inner disk is replaced by an ADAF for LLAGN.  Such a 
model provides a more general description of the accretion flow, which accommodates the diverse 
characteristics of LLAGN over a wide wavelength regime. 

\subsection{Constraints to the Viscosity Parameter and Magnetic Fields}
It has been shown that evaporation caused by the disk corona interaction is a promising mechanism to 
facilitate the truncation of an optically thick and geometrically thin disk in LLAGN, providing an 
understanding of the distinctive spectral states of HLAGN and LLAGN.  Within this framework, it may be possible 
to constrain the viscosity and magnetic field parameters within an $\alpha$, $\beta$ disk prescription. 
From Eq.(\ref{tr-radius}) it can be seen that the truncation radius strongly depends on the magnetic 
field while it is relatively insensitive to the viscosity parameter in cases where the thin disk recedes 
to large distances.  This provides a probe for exploring the magnetic field effects from observations via the 
derived truncation radius and mass accretion rates inferred from spectral fits.  For systems approaching a 
transition to a soft state, the relation between accretion rate and truncation radius deviates from 
Eq.(\ref{tr-radius}) and is dependent on the viscosity (see Fig.1).  In particular, at the transition, the 
accretion rate ($\dot m_{\rm max} $, see Eq.(\ref{mdot-max})) strongly depends on the viscosity parameter. However, 
the influence of the magnetic field on $\dot m_{\rm max} $ is limited to within a factor of 1.33 from the 
non magnetic case ($\beta=1$) to an equipartition value field ($\beta=0.5$), implying that the transition 
rate is determined primarily by the viscosity.  On the other hand, the truncation radius is much more 
sensitive to the magnetic field effects than to the viscosity (see Eq.(\ref{tr-min})). Hence, the observed 
transition luminosity (Eddington ratio) can be used to constrain the viscosity parameter and the inferred 
truncation radius can probe the relative importance of magnetic effects in the truncation process. 

\section{Conclusion}
We have studied the disk truncation model of LLAGN within the framework of the disk evaporation model. It 
is shown that the disk is truncated at mass accretion rates below 0.03 times the Eddington value for a 
standard viscosity parameter of $\alpha=0.3$.  For increasing values of $\alpha$, the critical accretion 
rate for disk truncation increases.  The inclusion of magnetic energy densities in the disk do not 
significantly modify this critical accretion rate, but result in truncation of the geometrically thin, 
optically thick disk at smaller radii closer to the central black hole. The model predicts that 
accretion flows in LLAGN take place via an inner ADAF and an outer geometrically thin disk. The truncated 
disk recedes outward with decreasing accretion rate and its radius is given as a function of accretion rate 
and magnetic fields based on approximate fits to the detailed numerical calculations of the disk corona 
interaction. We show that the truncated disk in LLAGN can emit in the infrared to optical band, depending 
on the specific mass of the black hole and the mass accretion rate.  Adopting a model of an inner ADAF 
connected to an outer truncated disk, we calculate a sample spectrum of a LLAGN.  The model predictions are 
qualitatively compared with the spectral fits to individual objects, and it is found that the evaporation 
model can provide a consistent framework for their interpretation.  This suggests that the truncation of a 
thin disk in LLAGN, a consequence of the disk corona interaction, is similar to the case in BHXRBs extending 
the accretion flow analogy between these two types of systems.
 
\acknowledgments
R.E.T. acknowledges support from the Theoretical Institute for Advanced Research in Astrophysics 
in the Academia Sinica Institute of Astronomy \& Astrophysics. Financial support for this work is provided by the National Natural Science Foundation of China (grants 11033007 and 11173029) and by the National Basic Research Program of China-973 Program 2009CB824800.

\vfil\eject

\begin{figure}
\plotone{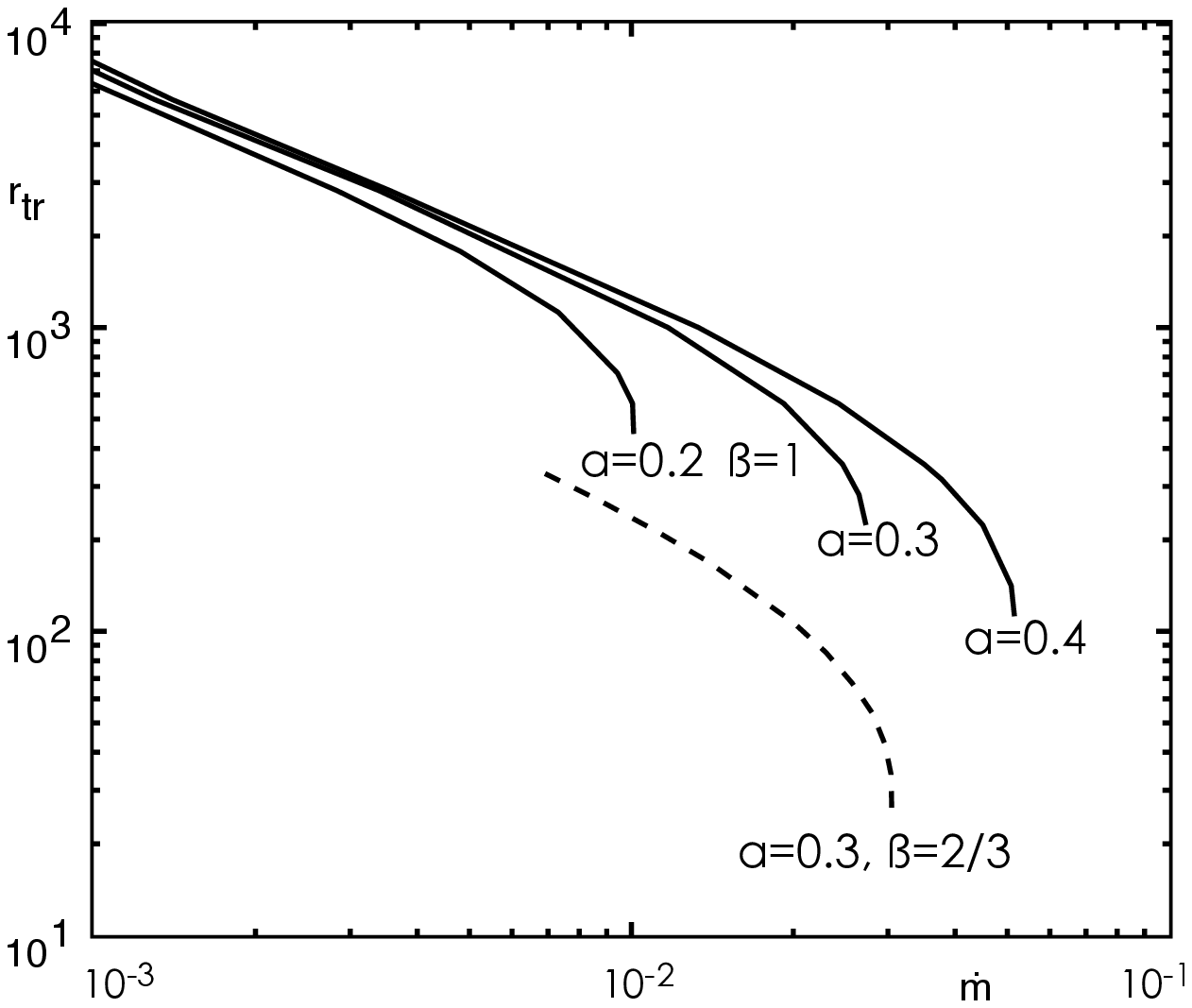}
\caption{\label{f:fig1} The variation of the truncation radius with respect to the mass accretion rate 
for specific values of the viscosity parameter $\alpha$ and magnetic parameter $\beta$ as predicted 
by the disk evaporation model.  Solid lines are for the case without magnetic effects, $\beta=1$ for a 
range in the viscosity parameter. 
The transition from the hard state to soft state takes place at lower accretion rates for a lower value 
of $\alpha$. The effect of magnetic field is shown by the dashed lines for  $\beta=2/3$ and  $\alpha$=0.3.
It can be seen that disk truncation occurs at a smaller radius with the inclusion of magnetic effects, however,
the transition accretion rate is little affected.}
\end{figure}

\begin{figure}
\plotone{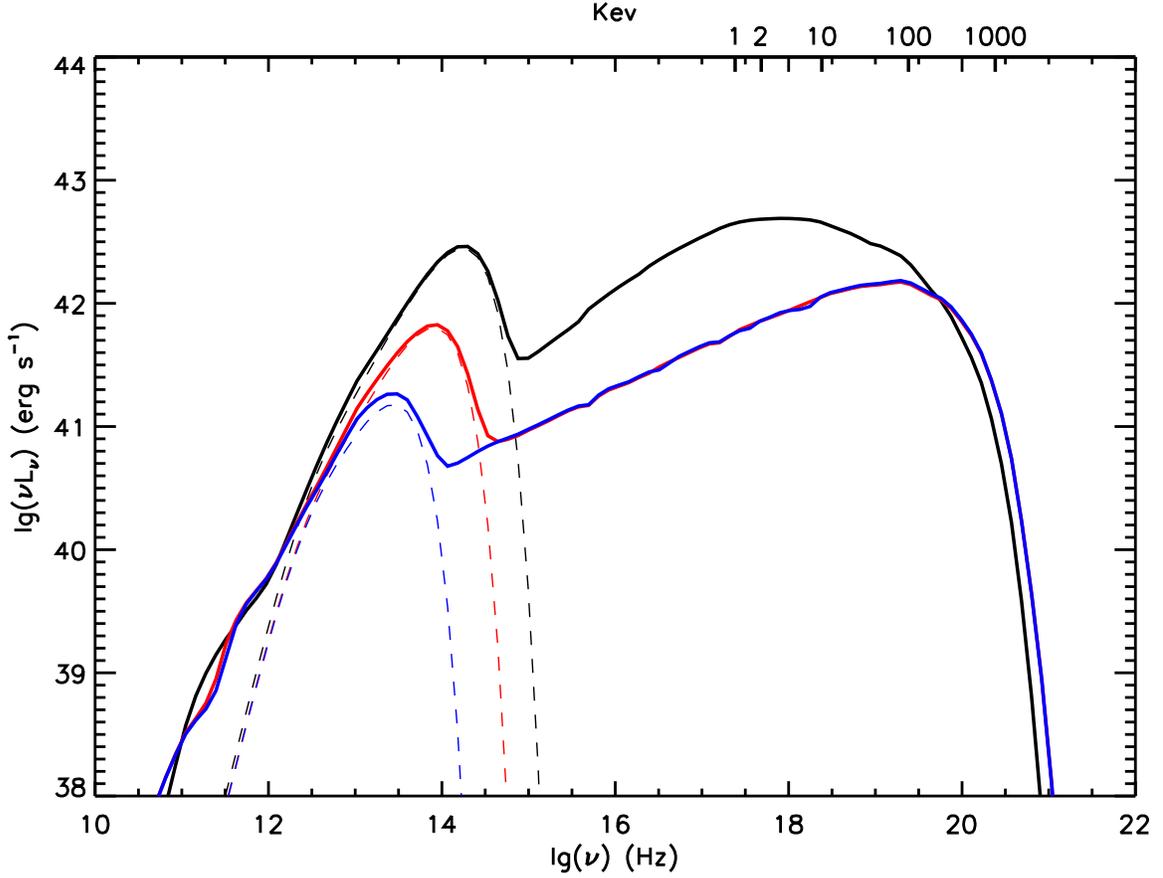}
\caption{\label{f:spectra} The radiation spectra from the inner ADAF and truncated disk around a SMBH of 
$m=10^{8}$ assuming $\alpha=0.3$.  The black solid curve corresponds to $\dot m=0.02$ and $\beta=2/3$,  
where a thin disk is truncated by evaporation at a distance of $r_{\rm tr}=107$. The red solid curve is 
for $\dot m=0.01$ and  $\beta=2/3$, with  truncation radius of  $r_{\rm tr}=236$ determined by the 
evaporation model.  The blue solid curve is for $\dot m=0.01$ and  $\beta=1$,  with truncation radius of 
$r_{\rm tr}=1000$.  The dashed lines show the radiation contributed by the truncated disk. It can be seen 
that the emission from a truncated disk forms a red bump.  The frequency corresponding to the
bump depends on the accretion rate and strength of the magnetic field.}
\end{figure}
\end{document}